\documentclass[preprintnumbers, pra, showpacs, floatfix,twocolumn,
preprintnumbers, letterpaper, superscriptaddress]{revtex4}
\usepackage{amsfonts}
\usepackage{amsmath}
\usepackage{amssymb,epsf}
\usepackage{latexsym}
\usepackage{graphicx,epsfig}
\usepackage{amssymb}
\usepackage{subfigure}
\usepackage[colorlinks=true,citecolor=blue,linkcolor=blue,urlcolor=black]{hyperref}
\usepackage{epstopdf}
\usepackage{color}

\def\be{\begin{equation}}
\def\ee{\end{equation}}
\def\ba{\begin{eqnarray}}
\def\ea{\end{eqnarray}}
\def\la{\langle}
\def\ra{\rangle}

\begin{document}
\title{Noisy Toric code and random bond Ising model: \\
The error threshold in a dual picture}
\author{Mohammad Hossein Zarei}
\email{mzarei92@shirazu.ac.ir}
\affiliation{Physics Department, College of Sciences, Shiraz University, Shiraz 71454, Iran}
\author{Abolfazl Ramezanpour}
\email{aramezanpour@gmail.com}
\affiliation{Physics Department, College of Sciences, Shiraz University, Shiraz 71454, Iran}
\affiliation{Leiden Academic Centre for Drug Research, Faculty of Mathematics and Natural Sciences, Leiden University, Leiden, The Netherlands}
\begin{abstract}
It is known that noisy topological quantum codes are related to random bond Ising models where the order-disorder phase transition in the classical model is mapped to the error threshold of the corresponding topological code. On the other hand, there is a dual mapping between classical spin models and quantum Calderbank-Shor-Stean (CSS) states where the partition function of a classical model defined on a hypergraph $H$ is written as an inner product of a product state and a CSS state on dual hypergraph $\tilde{H}$. It is then interesting to see what is the interpretation of the classical phase transition in the random bond Ising model within the framework of the above duality and whether such an interpretation has any connection to the error threshold of the corresponding topological CSS code. In this paper, we consider the above duality relation specifically for a two-dimensional random bond Ising model. We show that the order parameter of this classical system is mapped to a coherence order parameter in a noisy Toric code model. In particular, a quantum phase transition from a coherent phase to a non-coherent phase occurs when the initial coherent state is affected by two sequences of bit-flip quantum channels where a quenched disorder is induced by measurement of the errors after the first channel. On the other hand, the above transition is directly related to error threshold of the Toric code model. Accordingly, and since the noisy process can be applied to other topological CSS states, we conclude that the dual correspondence can also provide a useful tool for the study of error thresholds in different topological CSS codes.
\end{abstract}
\maketitle
\section{Introduction}
It has been shown that quantum information theory has many fascinating connections to statistical physics which have led to several cross-fertilizations between these two important fields of physics. During past decades different problems including quantum algorithms for evaluation of partition functions or simulation of classical spin models \cite{Geraci2008, Lidar1997, Geraci2010, algor, durmar}, mathematical mappings from quantum systems to classical statistical models \cite{Somma2007, eis17, castel2005, gemma} and applications of percolation theory in quantum error correcting codes \cite{vodola} have been developed.  Of the most interesting connections is a mapping from partition function of classical spin models to quantum stabilizer states \cite{Nest2007}. Such mappings have recently been simplified by a duality relation based on hypergraphs \cite{zare18} where partition function of a classical spin model defined on a hypergraph $H$ can be written as an inner product of a product state and a CSS state defined on the dual hypergraph $\tilde{H}$. Using the above dual correspondence, concepts from measurement based quantum computation have been used for considering complexity of classical spin models \cite{Bravyi2007,Bombin2008}. In particular, concept of completeness in statistical models \cite{Nest2008,Vahid2012b,Cuevas2009,xu,Vahid2012,yahya} and recently universal models \cite{science,cub} have emerged in statistical physics which has opened an important window toward a unification in statistical mechanical models.

On the other hand, one of important and well-established problems in statistical mechanics is the existence of thermal phase transition in classical spin models \cite{pathr}. In this respect and using the dual correspondence between classical spin models and quantum CSS states, one can expect to find an interpretation of phase transition in classical models for quantum CSS states. For example, it has recently been shown that thermal phase transitions in classical spin models are mapped to topological phase transitions in the corresponding CSS models \cite{zarei19}. Furthermore, the above mapping has led to insights about noisy topological CSS states where thermal phase transition in a ferromagnetic spin model is mapped to a critical stability against bit-flip noise in the corresponding topological CSS code\cite{zare18}. In particular,  it has been shown that phase transition in two-dimensional (2D) Ising model is mapped to a coherence-noncoherence phase transition in the 2D Toric code model under a bit-flip noise \cite{zareimon}.

In addition to the dual correspondence, connection between thermal phase transition in classical spin models and noisy topological codes has also been considered in a different framework \cite{Dennis2002,Katzgraber2009,q1,q2,q3,q4,3dcolor}, where the error threshold of noisy topological codes is mapped to the transition point of the random bond Ising models on the Nishimori line \cite{nishi} . Accordingly, it seems that such problem should be related to the dual correspondence between classical spin models and quantum CSS states. If it is correct, it means that the same dual correspondence also relates the error threshold of a topological CSS state defined on a hypergraph $H$ with the phase transition point of a random bond Ising system defined on $\tilde{H}$.

In this paper, we start from the dual correspondence for the 2D random bond Ising model where the partition function is mapped to an inner product of the Toric code state and a product state in which random couplings are encoded. Then, we explore a mapping from this problem to a noisy Toric code model to find a relation to the error threshold problem. To this end, we introduce a quantum formalism for the order parameter of the random bond Ising model. Then, using a mapping from temperature to the probability of noise, we find an important interpretation for the order parameter in a noisy Toric code model. Specifically, we introduce a noisy process on a coherent state in the Toric code where two sequences of bit-flip channels are applied and a quenched disorder is induced by measurement of error syndromes after the first channel. We introduce an order parameter for determining the coherence of an initial state after applying noise, and we show that it is mapped to local magnetization in the random bond Ising model. Therefore, we find a coherence-noncoherence phase transition in the noisy Toric code model corresponding to ferromagnetic-paramagnetic phase transition in the random bond Ising model. Specifically, the phase diagram here is richer than the one that has been derived for ordinary Ising model \cite{zareimon}, in the sense that here the two quenched and annealed noise parameters determine the coherence phase of the model. On the other hand, we show that the above transition is also related to the error threshold of the Toric code model. Since this result is derived by using the dual correspondence and it can be applied to arbitrary random bond Ising systems, we conclude that the same dual correspondence connects the error threshold of topological CSS codes and the transition point of the dual random bond Ising systems. We give examples to show that this is very useful for studying the error threshold of different topological states. The reason is that by using hypergraphs we can easily find the classical spin model corresponding to any quantum CSS state.

The paper is organized as follows. In Sec.(\ref{sec1}) we briefly review the quantum formalism of the partition function for Ising model. In Sec.(\ref{sec2}), we introduce a quantum formalism for the order parameter of the random bond Ising model. In Sec.(\ref{sec3}), we give the main result of the paper where we introduce a mapping to Toric code model under a sequence of bit-flip noises. Specifically, we map the order parameter of the random-bond Ising model to a coherence parameter for the above noisy Toric code model. Finally, in Sec.(\ref{sec4}) we explain the connection of our results with the error threshold of the Toric code and we conclude that the duality correspondence between classical spin models and quantum CSS states is helpful for investigation of the error threshold of quantum CSS states.
\section{Quantum formalism for partition function of 2D Ising model}\label{sec1}
In this section, we give a brief review of the quantum formalism for partition function of the two-dimensional (2D) Ising model. To this end, consider a 2D square lattice where classical spins of $s=\pm 1$ live on vertices (vertex spins). The classical Hamiltonian is defined as follows:
\begin{equation}\label{ran}
H=-\sum_{\la i, j\ra} J_{ij}s_i s_j,
\end{equation}
where $\la i, j\ra$ refers to the nearest neighbors and $J_{ij}$ refers to the coupling constant corresponding to the interaction of  $\la i, j\ra$.
All thermodynamic properties of the above model can be derived from the partition function which is obtained by a summation of the Boltzmann weights corresponding to all spin configurations $\mathbf{s}$:
\begin{equation}\label{ranz}
\mathcal{Z}[\mathbf{J}]=\sum_{\mathbf{s}}e^{\beta \sum_{\la i, j\ra} J_{ij}s_i s_j},
\end{equation}
where $\beta$ is equal to $\frac{1}{k_B T}$ and $k_B$ refers to the Boltzmann constant.
\begin{figure}[t]
\centering
\includegraphics[width=8cm,height=4cm,angle=0]{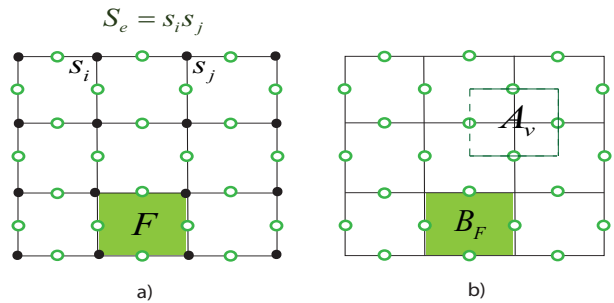}
\caption{(Color online) a) Ising spins $s_i$ denoted by black circles live on vertices of a square lattice. Edge spins $S_{ij}=s_i s_j$ denoted by green (light) circles are added on links of the lattice. Product of edge spins belonging to a face $F$ of the lattice should be equal to $1$. b) In a quantum formalism, each constraint corresponding to a face of the lattice is mapped to a face operator $B_F =\prod_{e\in \partial F}Z_e$. It is in fact a stabilizer of the Toric code. There are also another kind of stabilizers $A_v =\prod_{e\in v}X_e$ corresponding to each vertex of the lattice.}
\label{k4}
\end{figure}
Interestingly, the above partition function can be rewritten in a quantum formalism as follows.  We insert new spins $S_e$, (edge spins) denoted by green (light) circles, on the edges of the square lattice, as Fig.(\ref{k4}-a) shows. The value of each edge spin $S_e$ is defined as product of the vertex spins living on the two endpoints of the edge $e$, i.e., $S_e =s_i s_j$. Next, we replace the vertex spins in Eq.(\ref{ranz}) with edge spins $S_e$. However, we need also to consider the constraints between the edge spins. Specifically, as it is shown in Fig.(\ref{k4}-a), corresponding to each face $F$ of the lattice there is a constraint between edge spins in the form of $\prod_{e\in \partial F} S_e =1$ where $e\in \partial F$ refers to an edge around the face $F$. We apply such constraints by delta functions in Eq.(\ref{ranz}) and rewrite the partition function in the following form:
\begin{equation}\label{cons0}
\mathcal{Z}[\mathbf{J}]=\sum_{\mathbf{S}}e^{\beta \sum_{e} J_{e}S_e }\prod_{F} \delta(\prod_{e \in \partial F} S_e ,1).
\end{equation}
Here $\prod_F$ refers to all independent constraints corresponding to faces of the lattice and $J_e=J_{ij}$ where $i, j$ are the two endpoints of $e$. Furthermore, using the following form of the delta function $\delta(\prod_{e \in \partial F} S_e ,1)=\frac{1+\prod_{e \in \partial F} S_e}{2}$, we obtain:
\begin{equation}\label{cons}
\mathcal{Z}[\mathbf{J}]=\sum_{\mathbf{S}}e^{\beta \sum_{e} J_{e}S_e }\prod_{F} \frac{1+\prod_{e \in \partial F} S_e}{2}.
\end{equation}

Now, we are ready to introduce the quantum formalism of the partition function by replacing edge spins $S_e$ in Eq.(\ref{cons}) with Pauli operators $Z_e$. To this end, note that each edge spin $S_e$ takes two values $\pm1$ which can be considered as eigenvalues of a Pauli operator $Z_e$. Let $|0\ra$ and $|1\ra$ refer to eigenstates of the $Z$ operator with eigenvalues $+1$ and $-1$, respectively. It is simple to check that $\sum_{S_e} g(S_e) =2 \la +| g(Z_e)|+\ra$ where $g$ is an arbitrary function and $|+\ra =\frac{1}{\sqrt{2}}(|0\ra +|1\ra )$. Therefore, the partition function in Eq.(\ref{cons}) reads as follows:
\begin{equation}\label{qf}
\mathcal{Z}[\mathbf{J}]=2^M    ~^{\otimes M}\la +|e^{\beta \sum_{e} J_{e}Z_e }\prod_{F} \frac{1+\prod_{e \in \partial F} Z_e}{2}|+\ra ^{\otimes M},
\end{equation}
where $M$ refers to the number of edges of the lattice. The right hand side of the above equation is in fact an inner product of the following quantum states:
$$
|\alpha[\mathbf{J}] \ra =e^{\beta \sum_{e} J_{e}Z_e }|+\ra^{\otimes M}=\frac{1}{\sqrt{2}^M}\otimes_{e}(e^{\beta J_e }|0\ra +e^{-\beta J_e }|1\ra),
$$

\begin{equation}\label{inn}
|G\ra =\prod_{F} \frac{1+B_F}{2}|+\ra ^{\otimes M}.
\end{equation}
Here $B_F  =\prod_{e \in \partial F} Z_e$ which is called a face operator, see Fig.(\ref{k4}-b). Finally, the partition function is in the form of $\mathcal{Z}[\mathbf{J}]=2^M \la \alpha[\mathbf{J}] |G\ra$. While $|\alpha \ra$ is simply a product state containing information of the couplings of the Ising model, $|G\ra$ is an entangled state which encodes interaction pattern of the Ising model. Specifically, $|G\ra$ is a Toric code state defined on qubits that live on the edges of the lattice. Note that since $B_F (1+B_F)=(1+B_F )$, it is clear that state $|G\ra$ is stabilized by $B_F$. On the other hand, corresponding to each vertex of the lattice or each face of the dual lattice, there is a vertex operator in the form of $A_v =\prod_{e \in v}X_e$, see Fig.(\ref{k4}-b). Here $X_e$ refers to the Pauli operator $\sigma_x$ which is applied to the edge qubit on edge $e$ emanating to vertex $v$, see Fig.(\ref{k4}-b). Since each vertex operator $A_v$ has zero or two common qubits with $B_F$ operators, it is concluded that $[A_v ,B_F]=0$. Finally, if we apply $A_v$ to state $|G\ra =\prod_{F} \frac{1+B_F}{2}|+\ra ^{\otimes M}$ it will pass from $\prod_{F} \frac{1+B_F}{2}$ and since $A_v |+\ra ^M =|+\ra^M$, it is concluded that the $A_v$'s are also stabilizers of $|G\ra$. Therefore, state $|G\ra$ can also be written in terms of vertex operators in the following form:
\begin{equation}\label{innx}
|G\ra =\prod_{v} \frac{1+A_v}{2}|0\ra ^{\otimes M}.
\end{equation}
 In this way, the state $|G\ra$ is stabilized by vertex and face operators $A_v$ and $B_F$ which are exactly stabilizers of Toric code state as defined in \cite{Kitaev2003}. Consequently, the partition function of 2D Ising model has been mapped to inner product of a product state and the Toric code state where couplings of the Ising model are encoded in the product state and interaction pattern of the Ising model is encoded in the Toric code state.
\begin{figure}[t]
\centering
\includegraphics[width=8cm,height=4cm,angle=0]{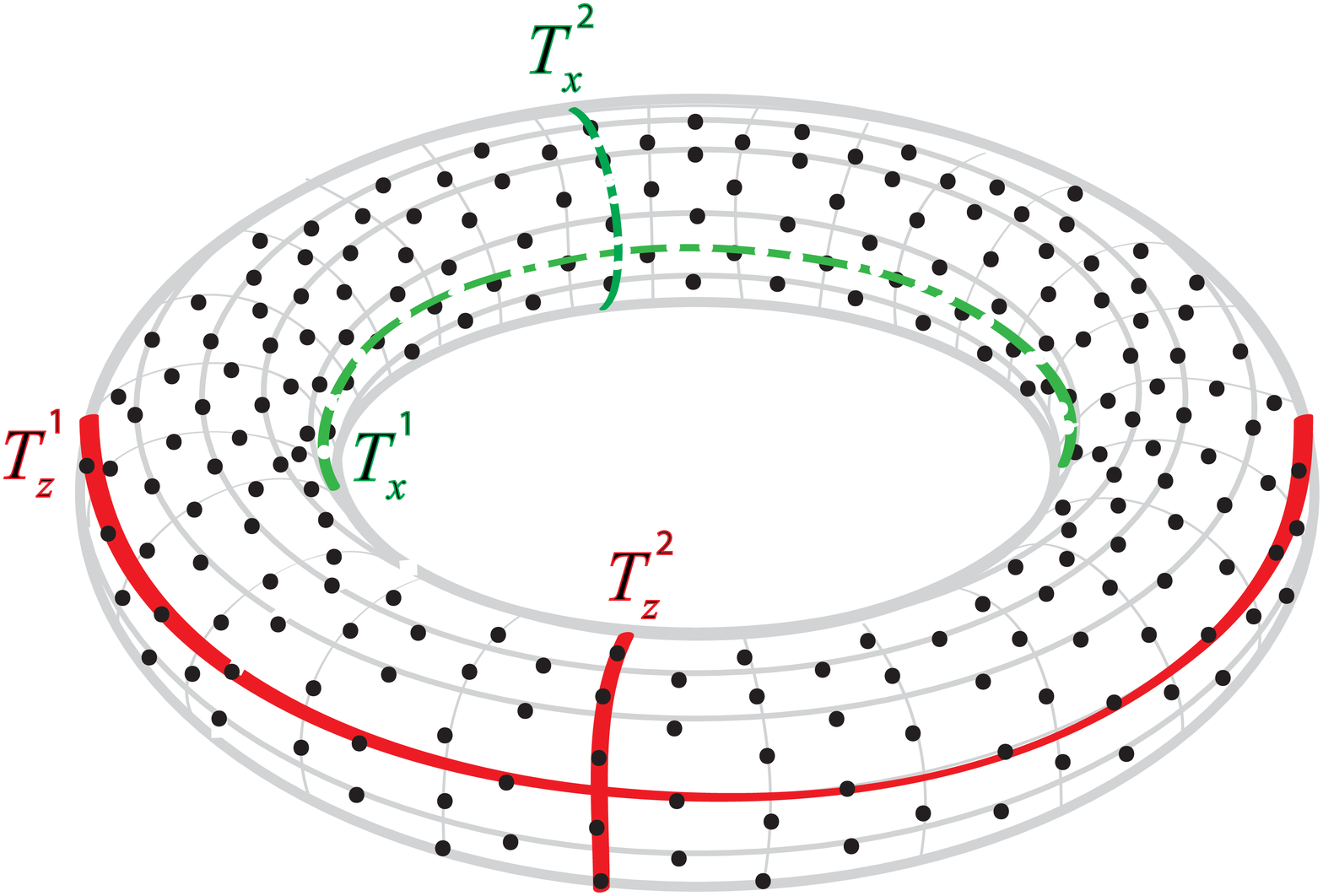}
\caption{(Color online) Loop operators $T_z ^1$, $T_z ^2$, $T_x ^1$ and $T_x ^2$ correspond to non-trivial loops around the torus.}
\label{k1}
\end{figure}
In this way, 2D Ising model is mapped to the Toric code state as an entangled state with topological order which shows a natural robustness against local perturbation \cite{rob1,rob2,zare16,zareiprb}. In order to understand topological order of this state, let us consider the above mapping for a lattice with a periodic boundary condition, i.e., a torus. With such a topology, it is necessary to reconsider the constraints which have been considered in Eq.(\ref{cons0}). In fact, the constraints associated with faces of the lattice are not independent when we consider a periodic boundary condition. As a result, the product of all face operators in Toric code on a torus is equal to Identity operator. Furthermore, there are other constraints which should be added because of the topology of the torus. In particular, consider two non-trivial loops as shown in Fig.(\ref{k1}) around the torus in two different directions. Since $Z$-type loop operators, denoted by $T^1 _z$ and $T^2 _z$, corresponding to the above non-trivial loops, can not be constructed by a product of face operators, they should be considered as two independent constraints. Consequently, state $|G\ra$ includes both trivial and non-trivial loop operators. The role of non-trivial loop operators in $|G\ra$ is a property which can not be characterized by any local order parameter. As it is shown in Fig.(\ref{k1}), there are two non-trivial loops on the dual lattice and we can define two $X$-type operators corresponding to such loops, denoted by $T^1 _x$ and $T^2 _x$. Since these loop operators cross $Z$-type non-trivial loop operators once, they have an anti-commutation relation in the following form:
\begin{equation}\label{non}
\{T^1 _x , T^2 _z \}=0~~,~~\{T^1 _z , T^2 _x \}=0.
\end{equation}
Now, if we apply operators $T^1 _x$ and $T^2 _x$ to state $|G\ra$, we obtain the four-fold degenerate ground space of the Toric code:
\begin{equation}\label{deg}
|\psi_{\mu \nu}\ra =(T^1 _x)^\mu (T^2 _x )^\nu |G\ra,
\end{equation}
with $\mu , \nu =\{0,1\}$. Because of the anticommutation relation(\ref{non}), it is concluded that the expectation values of $T^1 _z$ and $T^2 _z$ are different for the above quantum states. Since $T^1 _x$ and $T^2 _x$ are non-local operators, it means that the above quantum states can be distinguished from each other only by non-local order parameters. This property, that there is no local order parameter which can distinguish the above states from each other, is in fact the result of topological order of Toric code states. We should emphasize that here we considered the periodic boundary condition in order to better explain the topological order in the Toric code. However, in the following, we will return to the open boundary condition which is more convenient for our calculations.
\begin{figure}[t]
\centering
\includegraphics[width=6cm,height=4cm,angle=0]{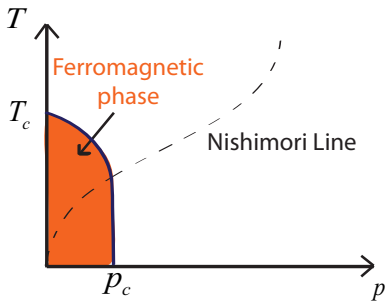}
\caption{(Color online) A schematic of phase diagram of random bond Ising model where ferromagnetic and paramagnetic phases are separated by a critical curve. There is also a Nishimori line where $p=\frac{e^{-2\beta J}}{1+e^{-2\beta J}}$. }
\label{schem}
\end{figure}
\section{Quantum formalism for the order parameter}\label{sec2}
2D Ising model is a well-known statistical mechanical model which can be used for describing the ferromagnetic-paramagnetic phase transition when all couplings in Eq.(\ref{ran}) are fixed to $J$. However, one can consider a quenched disorder for couplings to generate a random bond Ising model where each coupling might be $J$ with a probability of $1-p$ and might be $-J$ with a probability of $p$. The set of all edges $E$ then includes two subsets of $E_1$ and $E_2$ where $E_1 \cup E_2 =E$; all edge couplings in subset $E_1$ are fixed to $+J$ and the couplings in subset $E_2 $ are fixed to $-J$. It has been shown that such a model can still show a ferromagnetic-paramagnetic phase transition. Such phase transition can be characterized by a local order parameter which is the averaged magnetization $m$, see Fig.(\ref{schem}) as a schematic of phase diagram of random bond Ising model. 

In the previous section we gave a quantum formalism for the partition function of such a model. In this section we are going to find a quantum formalism for the magnetization $m$ in order to consider consequences of phase transition in 2D Ising model in the quantum side. Such a study has been done for the Ising model without quenched disorder where the above phase transition is mapped to a coherent/non-coherent phase transition in the noisy Toric code state \cite{zareimon}. However, here we are going to consider random-bond Ising model which has a richer phase diagram and the magnetization is a function of the temperature and the probability of disorder $p$.

We start with a quenched configuration of couplings to calculate the magnetization and then we calculate the mean value of this quantity for all realizations of the quenched disorder. For an arbitrary spin, for example $s_n$:
\begin{equation}\label{deg}
m_n[\mathbf{J}]=\la s_n \ra=\frac{1}{\mathcal{Z}[\mathbf{J}]}\sum_{\mathbf{s}}s_n e^{\beta \sum_{\la i,j\ra }J_{ij}s_i s_j },
\end{equation}
\begin{figure}[t]
\centering
\includegraphics[width=7cm,height=10cm,angle=0]{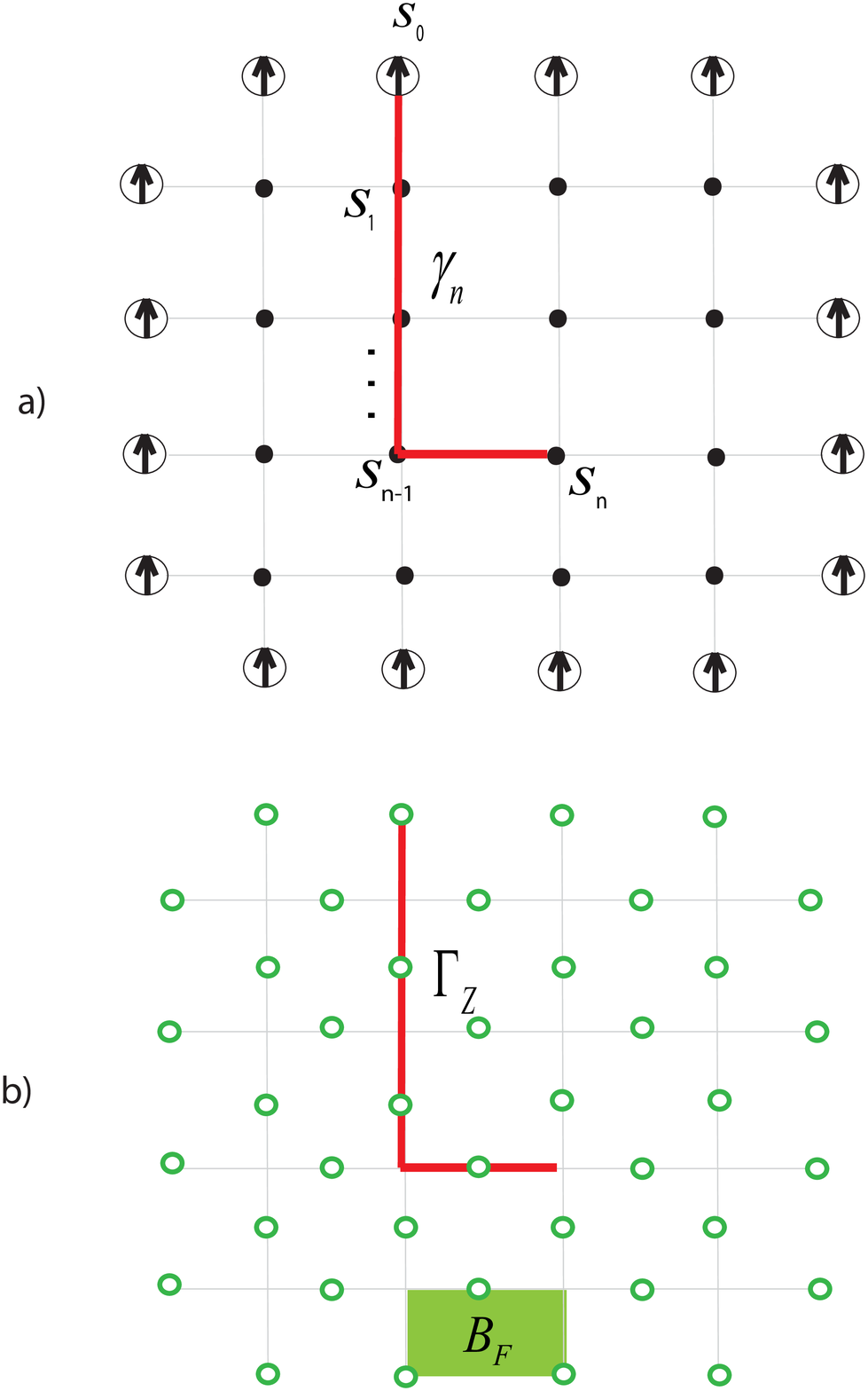}
\caption{(Color online) a) String $\gamma_n$ starts from a spin in the boundary and ends in a spin at site $n$. b) The Toric code state corresponding to the Ising model with the above boundary condition will have an open boundary where face operators in the boundary are three-local. To each string $\gamma_n$ we assign a $Z$-type operator $\Gamma_Z =\prod_{i \in \gamma_n }Z_i$.}
\label{string}
\end{figure}
where $J_{ij}=+J~(-J)$ when $i$ and $j$ are endpoints of an edge $e\in E_1  ~(E_2)$. In order to find a quantum formalism for the above relation, let us consider an open boundary condition for the model where spins of the boundary are fixed to $s_0 =+1$, see Fig.(\ref{string}). By such boundary condition, we consider a string called $\gamma_n$ which starts from a spin in the boundary and ends in spin $s_n$. Next, since $s_i ^2 =1$, we can rewrite $s_n$ as $s_n s_{n-1}s_{n-1} ... s_1 s_1 s_0$ and therefore, Eq.(\ref{deg}) is written in the following form:
\begin{equation}\label{deg2}
m_n[\mathbf{J}]=\frac{\sum_{\mathbf{s}}(s_n s_{n-1})(s_{n-1}s_{n-2})...( s_1 s_0) e^{\beta \sum_{\la i,j\ra }J_{ij}s_i s_j }}{\mathcal{Z}[\mathbf{J}]}.
\end{equation}
Next, we do the same process done in the previous section for finding the quantum formalism of the partition function. First we replace Ising interactions $s_i s_j$ with edge spins $S_e$ and then these spins are replaced with Pauli operators $Z_e$. Therefore, Eq.(\ref{deg2}) is written in the following quantum form:
\begin{equation}\label{deg3}
m_n[\mathbf{J}]=\frac{\la \alpha[\mathbf{J}] |\prod_{e\in \gamma_n}Z_e |G\ra}{\la \alpha[\mathbf{J}] |G\ra},
\end{equation}
where the operator $\Gamma_Z =\prod_{e\in \gamma_n} Z_e$ refers to a product of $Z$ operators corresponding to all edge qubits belonging to string $\gamma_n$. Furthermore, we note that since here we have an Ising model with open boundary, the state $|G\ra$ in the above relation is also the Toric code state defined on a lattice with open boundary condition. Thus, the face operators in the boundary should be a local operator in the form of a product of three $Z$ operators which is called a three-local operator, see Fig.(\ref{string}).  The operator $\Gamma_Z$ is in fact the quantum form of the product $(s_n s_{n-1})s_{n-1} ... s_1 (s_1 s_0)$ in Eq.(\ref{deg2}). Now we have a quantum formalism for the order parameter of the random bond Ising model with an arbitrary quenched configuration of disorder. We should emphasize that it is necessary to take the mean value $\overline{m_n[\mathbf{J}]}$ for different quenched configurations. Therefore, the mean value $m=(\sum_n \overline{m_n[\mathbf{J}]})/N$ will be a function of $\beta$ and $p$. In the next section, we find a physical meaning for this quantity where we show that it is related to a decoherence problem in the Toric code model.

\section{Mapping to the Toric code model under bit-flip channels}\label{sec3}
By now, we have only rewritten the partition function as well as the order parameter of the random bond Ising model in a quantum formalism. However, the above quantum formalism seems only a mathematical mapping and we need to introduce a physical meaning for the order parameter in the quantum side. In this section, we show that the order parameter can characterize coherence of a quantum state in the Toric code model which is under two sequences of bit-flip channels. To this end, first we need to introduce the bit-flip channel and consider its effect on the Toric code model. This section involves two subsections where in the first one we introduce the noisy Toric code and in the second one we come back to the quantum formalism of the order parameter and derive the main result of the paper.

\subsection{Toric code model and bit-flip noise}
As we briefly mentioned in the first section, Toric code model is defined by two kinds of stabilizer operators $B_F$ and $A_v$. Furthermore, the Toric code state $|G\ra$ is also a ground state of a Hamiltonian in the form of $H=-\sum_F B_F -\sum_v A_v$.  Now, suppose that the initial state is the same as the ground state $|G\ra$ and then a Pauli operator $X$ is applied to each qubit with probability $p$. Here we use the same notation of random bond Ising model because we are going to map these problems to each other. Therefore, the initial state will change to the following state after the bit-flip noise:
\begin{equation}\label{bit}
\rho=\sum_{\mathcal{E}} W_{\mathcal{E}}(p) \hat{\mathcal{E}}(X)|G\ra \la G| \hat{\mathcal{E}}(X),
\end{equation}
where $\mathcal{E}$ refers to a pattern of qubits which are affected by the noise and $\hat{\mathcal{E}}(X)$ refers to a product of $X$ operators which are applied to the qubits belonging to the error pattern $\mathcal{E}$. $W_{\mathcal{E}}(p)$ is the probability that an error patter $\mathcal{E}$ happens. If the number of qubits which have been affected by the noise is equal to $l$ and the total number of qubits is $M$, $W_{\mathcal{E}}(p)$ will be equal to $p^l (1-p)^{M-l}$.
\begin{figure}[t]
\centering
\includegraphics[width=6cm,height=6cm,angle=0]{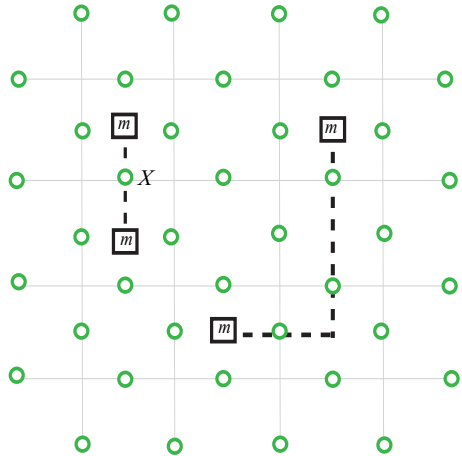}
\caption{(Color online) An $X$ operator on a qubit in the Toric code does not commute with two face operators which are neighbors of that qubit. It leads to two excitations in the neighboring faces. A string of $X$ operators also leads to two excitations in the endpoints of that string.}
\label{flux}
\end{figure}
It is useful to give a geometrical description for the effect of the bit-noise in the Toric code state. As it has been shown in Fig.(\ref{flux}), when an $X$ operator is applied to a qubit, it does not commute with two face operators which involve that qubit. Therefore, the system goes to an excited state which can be represented by two excitations, which are called flux anyons, in the corresponding faces. Consequently, the effect of the bit-flip operator on a qubit is geometrically described by a string in the dual lattice whose endpoints are flux anyons. In the same way, the effect of bit-flip noise on other qubits can be described by a pattern of strings whose endpoints show excitations. However, there can also be some closed strings (loops) where the pattern of effect of bit-flip operators is described by loops. Such patterns do not lead to an excitation and the Toric code state remains stable under such patterns. Consequently, the state $\rho$ is a mixture of excited states and the ground state of the Toric code model. In particular, one can compute the fidelity of $\rho$ with the ground state i. e. $F=\la G| \rho |G\ra$ which is also a measure of stability of the Toric code state against the bit-flip noise. This quantity will be in fact equal to total probability that bit-flip noise generates a pattern of loops in the dual lattice.

By the above geometrical description for noise patterns, it seems that some concepts like stability against noise might be related to some geometrical problems. Furthermore, other geometrical problems can emerge when we consider other initial states in Eq.(\ref{bit}). For example, we define a specific quantum state as a coherent superposition of the ground state and an excited state of the Toric code model, an excitation which is generated by applying $Z$ operators in the ground state. To this end, as it has been shown in Fig.(\ref{loop}), consider a Toric code model with an open boundary condition. Then consider a string $\gamma$ which starts from the boundary and ends to a vertex of the lattice, see Fig.(\ref{loop}). Then, we construct the string operator $\Gamma_Z =\prod_{e \in \gamma}Z_e$ corresponding to such a string. This operator commutes with all face operators. It also commutes with all vertex operators except the vertex operator corresponding to the vertex in the endpoint of the string. Therefore, $\Gamma_Z$ leads to an excitation called charge anyon in the Toric code model which is geometrically represented by a string in the lattice. Now, consider a coherent superposition from the ground state and the above excited state in the following form:
\begin{figure}[t]
\centering
\includegraphics[width=6cm,height=6cm,angle=0]{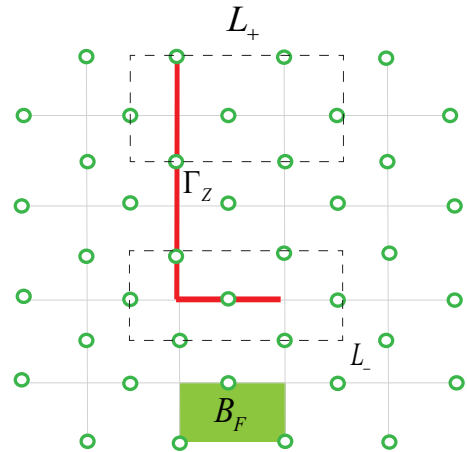}
\caption{(Color online)  Toric code can be defined on a lattice with an open boundary where face operators in the boundary are three-local. A string $\gamma$ which starts from the boundary leads to an excitation in the vertex in the end point of $\gamma$. There are two $X$-type loop operators $L_+$ and $L_-$ which cross the string $\gamma$ for even and odd number of times, respectively.}
\label{loop}
\end{figure}
\begin{equation}\label{str}
|\psi_+ \ra =\frac{1}{\sqrt{2}}(|G\ra +\Gamma_Z |G\ra ).
\end{equation}
Next, we consider the above state as an initial state which has been affected by a bit-flip noise. As we explained, the bit-flip noise can be described by patterns including open strings and loops. Similar to the Toric code state $|G\ra$, the effect of open strings on the state $|\psi_+ \ra$ leads to flux anyons. However, the effect of loops is different here. We divide the noise patterns that correspond to loops to two subsets, see Fig.(\ref{loop}). The first subset, denoted by $L_+$, contains loops that cross the string $\gamma$ an even number of times and the second subset, denoted by $L_-$, contains loops that cross the string $\gamma$ for an odd number of times. Since the $X$-type loop operators corresponding to the first subset commute with the operator $\Gamma_Z$, they preserve the state $|\psi_+ \ra$. On the other hand, the $X$-type loop operators corresponding to the second subset anti-commute with the $\Gamma_Z$ and therefore, they convert the state $|\psi _+\ra$ to a new state in the form of $|\psi_- \ra =\frac{1}{\sqrt{2}}(|G\ra -\Gamma_Z |G\ra$. It will be interesting to consider the effect of bit-flip noise only in a subspace generated by $|\psi_+ \ra$ and $| \psi_- \ra$. The initial state is a coherent state in this subspace and after bit-flip channel the quantum state in this subspace is converted to a completely mixed state. It has been shown that there is a quantum phase transition from coherent to the mixed phase in a critical value of the probability $p_{cr}$ \cite{zareimon}. Such a result has been derived by a mapping from the order parameter of the Ising model to an order parameter in Toric code state which characterizes the coherence in the above subspace. In the next subsection, we consider a different noise process on the $|\psi _+\ra $ where the coherence of the model is mapped to order parameter of random bond Ising model.
\subsection{Random bond Ising model and transition in coherence}
Let us start again with the initial state $|\psi_+ \ra$. Then, we consider two sequences of bit-flip channels of (\ref{bit}) where the probability of the noise is $p$ for the first channel and is $q$ for the second one. We also suppose that after the first quantum channel, a measurement of face operators $B_F$ is done in order to find what excitations have been generated by the noise. Therefore, after measurement we will have a quenched pattern of the error and then the second quantum channel is applied. Consequently, the final quantum state should be in the following form:
\begin{equation}\label{bit1}
\Phi[\boldsymbol\eta] =\sum_{\mathcal{E}} W_{\mathcal{E}}(q) \hat{\mathcal{E}}(X) \hat{\eta}(X) |\psi_+ \ra \la \psi _+| \hat{\eta}(X)\hat{\mathcal{E}}(X),
\end{equation}
where $\eta$ refers to the error pattern generated by the first channel. Similar to the previous subsection, here we give a geometrical description of the above state. The most important difference is that here we have a combination of two noise patterns related to operators $\hat{\mathcal{E}}(X)$ and $ \hat{\eta}(X)$. However, both the above operators are in the same type and both of them involve patterns of open strings and loops. Therefore, the initial state goes to its excitations if the combination of $\hat{\mathcal{E}}(X)$ and $ \hat{\eta}(X)$ corresponds to a pattern that includes open strings and it remains in the subspace of $|\psi _+\ra$ and $|\psi_ -\ra$ if the above combination corresponds to a pattern of loops. Furthermore, here we can also consider coherence in the above subspace where loops that cross the string $\gamma$ for an odd number of times lead to decoherence in the above subspace. In order to characterize the coherence in the above subspace, we define the following parameter as a measure of coherence:
\begin{equation}\label{coh}
O[\boldsymbol\eta]=\frac{W_+[\boldsymbol\eta] -W_-[\boldsymbol\eta]}{W[\boldsymbol\eta]}.
\end{equation}
Where $W_+[\boldsymbol\eta]$ is  the fidelity of state $\Phi[\boldsymbol\eta]$ with $|\psi_+ \ra$ in the form of $W_+[\boldsymbol\eta] =\la \psi_+ |\Phi[\boldsymbol\eta] |\psi_+ \ra$. Similarly, $W_-[\boldsymbol\eta]$ is the fidelity of $\Phi[\boldsymbol\eta]$ and $|\psi_- \ra$ which is $W_-[\boldsymbol\eta] =\la \psi_- |\Phi[\boldsymbol\eta] |\psi_- \ra$ and $W[\boldsymbol\eta]=W_+[\boldsymbol\eta] + W_-[\boldsymbol\eta]$. By a geometrical description, $W_+[\boldsymbol\eta]$ is interpreted as a probability that the combination of errors $\hat{\mathcal{E}}(X)$ and $ \hat{\eta}(X)$ generates loop configurations which cross string $\gamma$ an even number of times and $W_-[\boldsymbol\eta]$ is the probability that the above combination generates loop configurations which cross  $\gamma$ an odd number of times. We should also emphasize that the measurement that we have done after the first channel can lead to different patterns of $\hat{\eta}(X)$ with the corresponding probability. However, we have defined the coherent parameter $O[\boldsymbol\eta]$ for one specific quenched pattern, therefore one should take the mean value $O(p,q)=\overline{O[\boldsymbol\eta]}$ for different realizations of $\hat{\eta}(X)$. Since the probability of generating $\hat{\eta}(X)$ is a function of $p$, the mean value of the coherent parameter is a function of $p$ and $q$. Interestingly, we shall see that the coherence parameter $O(p,q)$ is mapped to the order parameter of random bond Ising model. To show this, consider the quantum form of the local magnetization:

\begin{equation}\label{deg4}
m_n[\mathbf{J}]=\frac{\la \alpha[\mathbf{J}] |\prod_{e\in \gamma_n}Z_e |G\ra}{\la \alpha[\mathbf{J}] |G\ra}.
\end{equation}
In this equation, for a quenched configuration of couplings $\mathbf{J}$, there is a product state $|\alpha[\mathbf{J}] \ra =\frac{1}{\sqrt{2}^M}\otimes_{e}(e^{\beta J_e }|0\ra +e^{-\beta J_e}|1\ra)$, where $J_e =+J$ for $e \in E_1$ and $J_e =-J$ for $e \in E_2$. Such a quenched configuration of disorder in couplings can be mapped to a quenched pattern of errors in the Toric code which was denoted by $\eta$. Then, we do a change of variable from $\beta J$ to a parameter $q=\frac{e^{-2\beta J}}{1+e^{-2\beta J}}$. Since $\beta J$ is a quantity between zero to infinity, it is concluded that $0\leq q\leq \frac{1}{2}$. by such a change of variable the state $(e^{\beta J }|0\ra +e^{-\beta J}|1\ra)$ for $e\in E_1$ is converted to $e^{\beta J}~(1+\frac{q}{1-q}X)|0\ra$ and  the state $(e^{-\beta J }|0\ra +e^{+\beta J}|1\ra)$ for $e\in E_2$ is converted to $e^{\beta J}~(X+\frac{q}{1-q}1)|0\ra=e^{\beta J}~X(1+\frac{q}{1-q}X)|0\ra$. Therefore, the state $|\alpha \ra$ is written in the following form:
 $$
  |\alpha[\mathbf{J}] \ra=e^{M\beta J}\otimes_{e}(1+\frac{q}{1-q}X_e) \otimes_{e\in E_2}X_e |0\ra ^{\otimes M}
  $$
 \begin{equation}\label{e2}
=\frac{e^{M\beta J}}{(1-q)^{M}}  \otimes_{e}((1-q)I+q X_e) \otimes_{e\in E_2}X_e |0\ra ^{\otimes M}.
 \end{equation}
Next, consider the term $\otimes_{e}((1-q)I+q X_e)$ in the above equation. If we expand this product, it generates different patterns of $X$ operators where an $X$ operator appears with probability $q$. Therefore, we have:
\begin{equation}
\otimes_{e}((1-q)I+q X_e)=\sum_{\mathcal{E}} W_{\mathcal{E}}(q) \hat{\mathcal{E}}(X),
\end{equation}
where $W_{\mathcal{E}}(q)$ is the probability of error pattern $\mathcal{E}$ which is defined in Eq.(\ref{bit}). On the other hand, the other factor $\otimes_{e\in E_2}X_e$ in Eq.(\ref{e2})) is also a quenched pattern of $\eta$ where each qubit has been affected by an $X$ operator with probability $p$. In this way, the state $|\alpha \ra$ includes a superposition of all error patterns generated by the two sequences of bit-flip channels. The quenched error after the first channel is generated by a measurement of the face operators.

Now, we are ready to consider the inner product $\la \alpha[\mathbf{J}] |\prod_{e\in \gamma_n}Z_e |G\ra$ in Eq.(\ref{deg4}). As we explained in Sec.(\ref{sec1}), since the $A_v$ operators can be represented by loops in the dual lattice, the state $|G\ra=\prod_{v}(1+A_v)|0\ra^{\otimes M}$ includes a superposition of $X$-type loop operators. We divide again all possible loop configurations to two subsets. The first subset called $L_+$ includes loop configurations which cross the string $\gamma_n$ an even number of times and the second subset called $L_-$ includes loop configurations which cross the string $\gamma_n$ for an odd number of times. Therefore, the state $\prod_{e\in \gamma_n}Z_e |G\ra$ will be a superposition of the loop configurations in $L_+$ with a positive weight and the loop configurations in $L_-$ with a negative weight:
\begin{equation}
\prod_{e\in \gamma_n}Z_e |G\ra =|G_+\ra -|G_-\ra.
\end{equation}
Consequently, the inner product $\la \alpha[\mathbf{J}] |\prod_{e\in \gamma_n}Z_e |G\ra$ is equal to $\la \alpha[\mathbf{J}] |G_+\ra -\la \alpha[\mathbf{J}]|G_-\ra$. In this way, we have an inner product of $| \alpha[\mathbf{J}]\ra$, as a superposition of all possible error patterns with the corresponding probabilities, with $|G_+\ra$ and $|G_-\ra$, as two kinds of loop configurations. Therefore, up to an irrelevant factor, $\la \alpha[\mathbf{J}] |G_+\ra$ ($\la \alpha[\mathbf{J}] |G_-\ra$) will be equal to probability that the bit flip noise generates loop configurations which cross the string $\gamma_n$ an even (odd) number of times which is the same as $W_+[\mathbf{J}]$ ($W_-[\mathbf{J}]$).  Furthermore, it is clear that denominator in Eq.(\ref{deg4}) is also equal to $W_+[\mathbf{J}] +W_-[\mathbf{J}]$ up to an irrelevant factor which is removed from denominator and nominator. Therefore, the order parameter of the random bond Ising model will be in the form of $m_n[\mathbf{J}]=\frac{W_+[\mathbf{J}] -W_-[\mathbf{J}]}{W_+[\mathbf{J}] +W_-[\mathbf{J}]}$ which is the same as definition of the coherence parameter in Eq.(\ref{coh}). Consequently, a ferromagnetic paramagnetic phase transition in the 2D random bond Ising model is exactly mapped to a transition from a coherent to a non-coherent phase in the noisy Toric code model. It is also important to compare our result with the result which has already been derived in \cite{zareimon}, where a similar phase transition has been found. We should emphasize that here we have a richer phase diagram with the two quenched and annealed noise parameters which determine the coherence phase of the model. Therefore, there is a coherence region in the phase diagram with a non-zero coherence order parameter. We will return to this point in the conclusion in order to show it can lead to different insights on the noisy Toric code model. 

\section{Connection to error correction in the Toric code}\label{sec4}
In this section, we want to point out an interesting connection between our problem and quantum error correction in the Toric code. In particular, in \cite{Dennis2002} authors have considered Toric code model defined on a torus under bit-flip noise where, after measuring error syndromes, errors are corrected by an active process. To this end, one has to apply a string of $X$ operators between two syndromes to generate a loop to correct the error. However, since the Toric code has been defined on a torus, there is a probability that the string applied between two syndromes generates a non-trivial loop around the torus. Since such a non-trivial loop leads to an error in the code space, the Toric code will be non-correctable under such a situation. Therefore, the probability of the noise $p$ should be below a threshold such that the probability of generating a non-trivial loop goes to zero when the lattice size goes to infinity. It technically means the quenched error syndrome and the string of correction are in the same homological class. In the above paper, the authors have shown that the probability of being in the same homological class is mapped to the partition function of the random bond Ising model. Therefore, error threshold of the Toric code is mapped to the critical point of random bond Ising model on Nishimori line where $p=\frac{e^{-2 \beta J}}{1+e^{-2 \beta J}}$. In this respect, it seems that there are similarities with our problem where we have also a mapping to random bond Ising model and $p=q$ is in fact the Nishimori line in our process.

In order to clarify the above connection, let us come back to Eq.(\ref{bit1}) and replace the initial state with the ground state of the Toric code model defined on a torus:
\begin{equation}\label{}
\Phi^*[\boldsymbol\eta] =\sum_{\mathcal{E}} W_{\mathcal{E}}(q) \hat{\mathcal{E}}(X) \hat{\eta}(X) |G \ra \la G| \hat{\eta}(X)\hat{\mathcal{E}}(X).
\end{equation}
In fact, we will have two sequences of bit-flip channels on the Toric code state with a measurement of syndromes after the first channel. In order to be in the Nishimori line, we suppose that for both error patterns $\hat{\mathcal{E}}$ and $\hat{\eta}$, probability of single-qubit noise is equal to $p$. Then, we consider the fidelity of the final state with the initial state in the form of $F^*[\boldsymbol\eta]=\la G|\Phi ^*[\boldsymbol\eta]|G\ra$. Using geometrical interpretation of error patterns, it is concluded that $F^*$ is equal to total probability that error pattern $\hat{\mathcal{E}}$ generates loop configurations when it is added to quenched pattern $\hat{\eta}$. It is indeed equal to the probability that error patterns $\hat{\mathcal{E}}$ and $\hat{\eta}$ are in the same homological class. On the other hand, as it has been shown in \cite{Dennis2002}, this quantity is directly related to correctability of the Toric code under bit-flip noise in the sense that a singularity in this quantity reveals a transition in correctability. Interestingly, $F^*$ is also proportional to partition function of random bond Ising model. To check this, one can start with quantum formalism of the partition function $\mathcal{Z}[\mathbf{J}]=2^M \la \alpha [\mathbf{J}]|G\ra$. Then, similar to the procedure for the order parameter, and by a change of variable in $|\alpha [\mathbf{J}] \ra$ in the form of $q=\frac{e^{-2 \beta J}}{1+e^{-2 \beta J}}$, one can show that $\mathcal{Z}[\mathbf{J}]$ is proportional to $F^*$. Therefore, it is concluded that ferromagnetic-paramagnetic transition in the random bond Ising model on the Nishimori line is mapped to a transition in the correctability of the Toric code.

 We note that we derive the above result by using dual correspondence between Toric code and 2D Ising model. Therefore, it seems that such a result can be obtained in other topological CSS codes which are related to classical spin models by the dual correspondence \cite{zare18}. It means that we have a useful tool for finding error thresholds in topological CSS codes; by using the hypergraph duality which was introduced in \cite{zare18}, one is able to find the classical spin model that is associated to an arbitrary topological CSS state. Then, one should numerically calculate the magnetization of the corresponding random bond spin model to find the transition point on the Nishimori line. In order to clarify this point, let us give some examples as follows.

 As it has been shown in \cite{zare18}, using duality of hypergraphs one can show that dual of color code on a D-colex (D dimensional color complex) is an Ising model with (D+1)-body interactions on a D-simplicial lattice. Consequently, in order to find the error threshold of the color code on a D-colex, it is enough to find the transition point in a random bond Ising model with (D+1)-body interaction on the D-simplicial lattice. Interestingly, this problem has recently been studied in \cite{3dcolor} where the authors have shown that the error threshold of the color code defined on a 3-colex is mapped to the transition point of an Ising system with four-body interactions defined on a 3-simplicial lattice which confirms our statement. For another example, in \cite{zare18} it has been shown that dual of the Toric code defined on an arbitrary graph in arbitrary dimension is an ordinary Ising model on the same graph. Therefore, transition point of random bond Ising models defined on an arbitrary graph leads to the error threshold of the Toric code on the same graph. We should emphasize that such correspondence has already been introduced for two and three dimensional examples while our result is held for any dimension.
 \begin{figure}[t]
\centering
\includegraphics[width=6cm,height=11cm,angle=0]{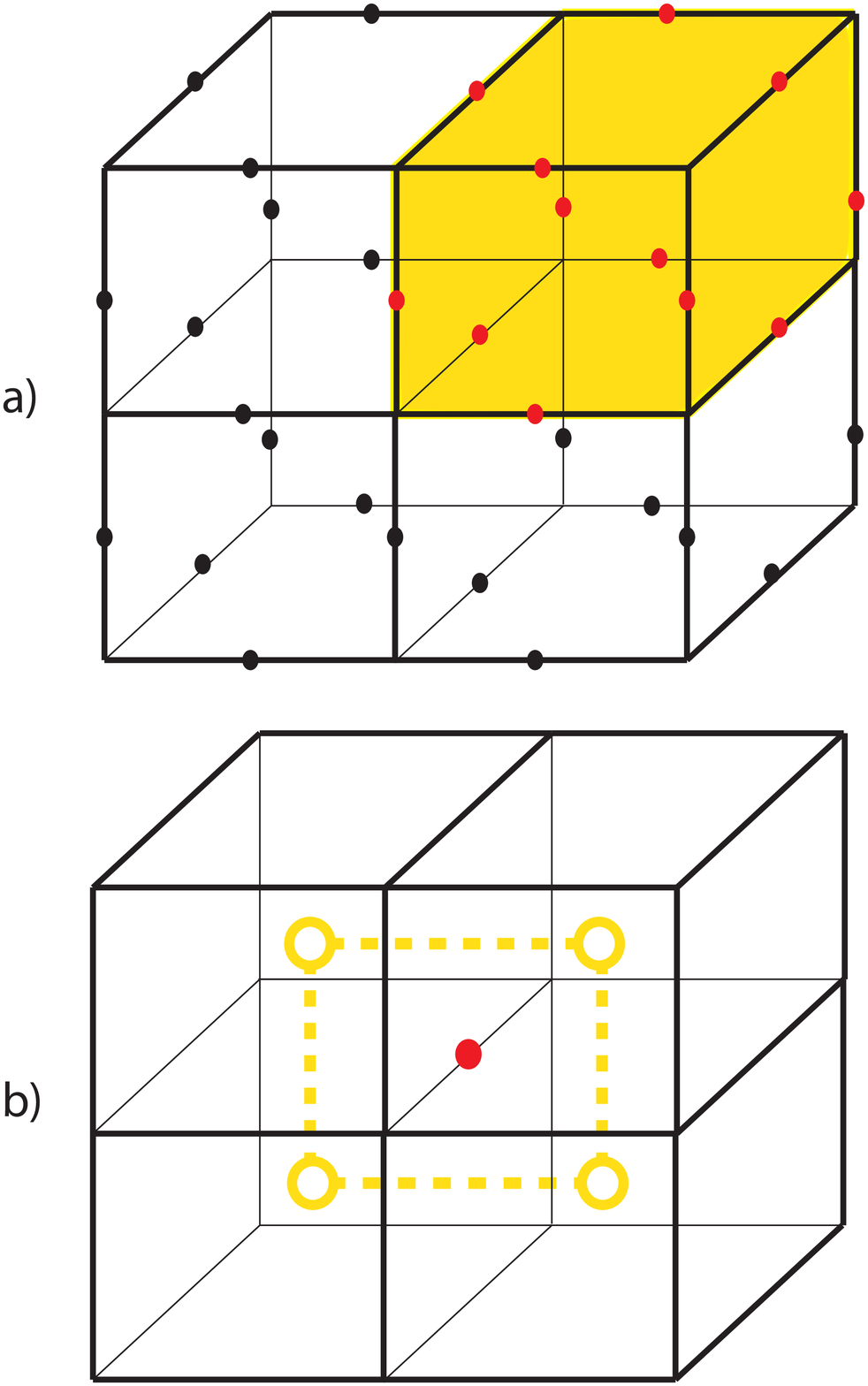}
\caption{(Color online) a) X-cube model is defined on a cubic lattice with qubits living on links of the lattice. Corresponding to each cell of the lattice denoted by yellow (light) color, an $X$ type stabilizer is defined by the product of $X$ operators for the 12 qubits belonging to the cubic cell. b) In a dual picture, one should insert spins in the center of each cell. Since each original qubit belongs to four cells of the lattice, each edge of the dual hypergraph involves four spins.  }
\label{xcub}
\end{figure}
 Furthermore, we note that the above duality is held for all CSS codes and it is not limited to archetypical examples including Toric codes and color codes. For example, a different quantum CSS code is $X$-cube model which has attracted much attention because of its more exotic properties than topological codes \cite{xcub0}. This code is defined on a cubic lattice with qubits living on links of the lattice, see Fig.(\ref{xcub}-a). $X$-type stabilizers of this code are defined by 12-body interaction terms corresponding to each cubic cell of the lattice. According to \cite{zare18}, we can define a hypergraph $H$ corresponding to this code where each cubic cell is equal to an edge of $H$ connecting the 12 qubits (vertices of $H$) of the cell. In order to find dual of the above hypergraph, we should exchange the role of vertices and edges in $H$. To this end, we insert new vertices in the center of each cubic cell of the lattice. The dual hypergraph $\tilde{H}$ is defined with these new vertices and the set of edges that correspond to the vertices of $H$. Since each vertex of $H$ (each link of the cubic lattice) is a member of four edges of $H$ (four cells of the cubic lattice), we conclude that each edge of $\tilde{H}$ has to involve four vertices corresponding to those cells, see Fig.(\ref{xcub}-b). Consequently, the dual hypergraph $\tilde{H}$ can be represented as a three dimensional lattice with vertices living on nodes of the lattice and each plaquette of the lattice corresponds to an edge of $\tilde{H}$. Then we can define a classical spin model corresponding to such a hypergraph where spins live on vertices and there is a four-body interaction for each plaquette of the 3D lattice. Finally, one can study the ferromagnetic phase transition in such a classical model with random couplings to find the error threshold of the $X$-cube model. We should emphasize that the concern of this study is not to find the error threshold for different topological CSS states. The aim of the above examples is to show how the hypergraph duality can be used to find the classical spin models corresponding to different topological CSS states.

\section{Discussion}
\begin{figure}[t]
\centering
\includegraphics[width=8cm,height=5cm,angle=0]{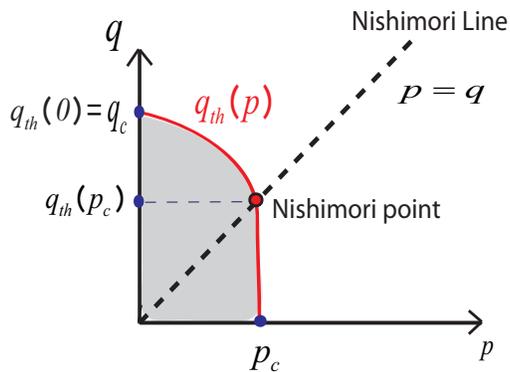}
\caption{(Color online) A schematic of phase diagram for the noisy Toric code model where transition point to the non-coherenct phase denoted by $q_{th}$ decreases by increasing quenched noise $p$ in a sense that corresponding to the transition curve, the $q_{th}$ is a decreasing function of $p$. The line of p=q refers to the Nishimori line and the transition point on the Nishimori line is called Nishimori point which determines the error threshold $q_{th}(p_c)$.  }
\label{phase}
\end{figure}

Although mapping the error threshold of topological codes to the phase transition of random bond Ising models has been established in the past decade, in this paper, we tried to reveal another aspect of those connections by using the quantum formalism of partition functions. First, we introduced a different interpretation for the transition point of the random bond Ising model in the quantum side as a quantum phase transition from a coherent to a non-coherent phase in a noisy Toric code model. Specifically, since there are two quenched and annealed noise parameters $(p,q)$ in the model, we find a richer phase diagram for the coherence phase of the model compared to the homogeneous model. In particular, the phase diagram, which has schematically been plotted in Fig.(\ref{phase}), shows that the quenched noise is more destructive to the coherence than the annealed noise, i.e. $p_c < q_c$. Moreover, the maximum level of the quenched noise that the coherence can tolerate is not very sensitive to the strength of the annealed noise up to the Nishimori point $q_{th}(p_c)$. On the other hand, the threshold value of the annealed noise goes abruptly to zero at the Nishimori point $p_c$, which coincides with the error correction threshold. 

Then, we showed that the problem of finding the error threshold of topological CSS codes is hidden in the quantum formalism of partition functions. Therefore, the duality correspondence between the partition functions of classical spin models and the quantum CSS states can be used to study the error threshold of topological CSS codes. As we explained in a few examples, by using the hypergraph duality correspondence, one would be able to find the classical spin model corresponding to an arbitrary CSS code. Therefore, our result provides a new tool for the study of error threshold in different CSS codes.

Finally, we would like to note that our results reveals the importance of studying different thermodynamic quantities in frame of quantum formalism of partition functions. While most studies had been done on partition functions, we showed that quantum formalism for the order parameter can also provide important insights about quantum stabilizer states. Therefore, it would be interesting to consider quantum formalism for other thermodynamic (well -known) quantities.

\section*{Acknowledgement}
We would like to thank A. Montakhab for his valuable comments on this paper.

\end{document}